# Chapter 4

# Management and Orchestration of Network Slices in 5G, Fog, Edge and Clouds

*Adel Nadjaran Toosi, Redowan Mahmud, Qinghua Chi and Rajkumar Buyya*

*Abstract*—Network slicing allows network operators to build multiple isolated virtual networks on a shared physical network to accommodate a wide variety of services and applications. With network slicing, service providers can provide a cost-efficient solution towards meeting diverse performance requirements of deployed applications and services. Despite slicing benefits, End-to-End orchestration and management of network slices is a challenging and complicated task. In this chapter, we intend to survey all the relevant aspects of network slicing, with the focus on networking technologies such as Software-defined networking (SDN) and Network Function Virtualization (NFV) in 5G, Fog/Edge and Cloud Computing platforms. To build the required background, this chapter begins with a brief overview of 5G, Fog/Edge and Cloud computing, and their interplay. Then we cover the 5G vision for network slicing and extend it to the Fog and Cloud computing through surveying the state-of-the-art slicing approaches in these platforms. We conclude the chapter by discussing future directions, analyzing gaps and trends towards the network slicing realization.

## 4.1 Introduction

The major digital transformation happening all around the world these days has introduced a wide variety of applications and services ranging from smart cities and vehicle to vehicle (V2V) communication to virtual reality (VR)/augmented reality (AR) and remote medical surgery. Design and implementation of a network that can simultaneously provide the essential connectivity and performance requirements of all these applications with a single set of network functions not only is massively complex but also is prohibitively expensive. The 5G Infrastructure Public-Private Partnership (5G-PPP) has identified various use case families of enhanced mobile broadband (eMBB), massive machine-type communications (mMTC), and ultra-reliable low latency communication (uRLLC) or Critical Communications that would simultaneously run and share the 5G physical multi-service network [1]. These applications essentially have very different Quality of Service (QoS) requirements and transmission characteristics. For instance, Video-on-demand streaming applications in eMMB category require very high bandwidth and transmitting a large amount of content. While mMTC applications, such as Internet of Things (IoT), typically have a multitude of low throughput devices. The differences between these use cases show that the *one-size-fits-all* approach of the traditional networks does not satisfy different requirements of all these vertical services.

      A cost-efficient solution towards meeting these requirements is slicing physical network into multiple isolated logical networks. Similar to server virtualization technology successfully used in Cloud computing era, network slicing intends to build a form of virtualization that partitions a shared physical network infrastructure into multiple end-to-end level logical networks allowing for traffic grouping and tenants' traffic isolation. Network slicing is considered as the critical enabler of the 5G network

where vertical service providers can flexibly deploy their applications and services based on the requirements of their service. In other words, network slicing provides a *Network-as-a-Service* (NaaS) model which allows service providers to build and set up their own networking infrastructure according to their demands and customize it for diverse and sophisticated scenarios.

Software Defined Networking (SDN) and Network Function Virtualization (NFV) can serve as building blocks of network slicing by facilitating network programmability and virtualization. Software-defined networking (SDN) is a promising approach to computer networking that separates the tightly coupled control and data planes of traditional networking devices. Thanks to this separation, SDN can provide a logically centralized view of the network in a single point of management to run network control functions. NFV is another trend in networking gaining momentum quickly with the aim of transferring network functions from proprietary hardware to software-based applications executing on general-purpose hardware. NFV intends to reduce the cost and increase the elasticity of network functions by building virtual network functions (VNFs) that are connected or chained together to build communication services.

With this in mind, in this chapter, we aim to review the state of the art literature on network slicing in 5G, Edge/Fog and Cloud computing, and identify the spectrum challenges and obstacles must be addressed to achieve the ultimate realization of this concept. We begin with a brief introduction of 5G, Edge/Fog, and Clouds and their interplay. Then, we outline the 5G vision for network slicing and identify a generic framework for 5G network slicing. We then review research and projects related to network slicing in Cloud computing context while we focus on SDN and NFV technologies. Further, we explore network slicing advance in emerging Fog and Edge Cloud computing. This leads us to identify the key unresolved challenges of network slicing within these platforms. Concerning this review, we discuss the Gaps and trends towards the realization of network slicing vision in Fog and Edge and Software-defined Cloud computing. Finally, we conclude the chapter.

Table 4.1 lists various acronyms and abbreviations referenced throughout the chapter.

Table 4.1- Acronyms and Abbreviations

| | |
|---|---|
| **V2V** | Vehicle to Vehicle |
| **VR** | Virtual Reality |
| **AR** | Augmented Reality |
| **5G** | 5th generation mobile networks or 5th generation wireless systems |
| **eMBB** | enhanced Mobile Broadband |
| **mMTC** | massive Machine-Type communications |
| **uRLLC** | ultra-Reliable Low Latency communication |
| **QoS** | Quality of Service |
| **IoT** | Internet of Things |
| **SDN** | Software Defined Networking |
| **NFV** | Network Function Virtualization |
| **VNF** | Virtualized Network Function |
| **MEC** | Mobile Edge Computing |
| **NaaS** | Network-as-a-Service |
| **NFaaS** | Network function as a Service |
| **SDC** | Software-defined Clouds |
| **VM** | Virtual Machine |
| **VPN** | Virtual Private Network |
| **NAT** | Network Address Translation |
| **SFC** | Service Function Chaining |
| **SLA** | Service Level Agreement |
| **CRAN** | Cloud Radio Access Network |
| **RRH** | Remote Radio Head |
| **BBU** | Baseband Unit |
| **FRAN** | Fog radio access network |

## 4.2  Background

**5G:** The renovation of telecommunications standards is a continuous process. Practicing this, 5th generation mobile network or 5th generation wireless system, commonly called 5G, has been proposed as the next telecommunications standards beyond the current 4G/IMT Advanced standards [3]. The wireless networking architecture of 5G follows 802.11ac IEEE wireless networking criterion and operates on millimeter wave bands. It can encapsulate Extremely high frequency (EHF) from 30 to 300 gigahertz (GHz) that ultimately offers higher data capacity and low latency communication [4].

The formalization of 5G is still in its early stage and expected to be mature by 2020. However, the main intentions of 5G include enabling Gbps data rate in a real network with least round trip latency and offering long-term communication among the large number of connected devices through high fault tolerant networking architecture [1]. Also, it targets to improve the energy usage both for the network and the connected devices. Moreover, it is anticipated that 5G will be more flexible, dynamic and manageable compared to the previous generations [5].

**Cloud Computing:** Cloud computing is expected to be an inseparable part of 5G services for providing an excellent backend for applications running on the accessing devices. During last decade, Cloud has evolved as a successful computing paradigm for delivering on-demand services over the Internet. The Cloud data centers adopted virtualization technology for efficient management of resources and services. Advances in server virtualization contributed to the cost-efficient management of computing resources in the Cloud data centers.

Recently, the virtualization notion in Cloud data centers, thanks to the advances in SDN and NFV, has extended to all resources including compute, storage, and networks which formed the concept of Software Defined Clouds (SDC) [2]. SDC aims to utilize the advances in areas of Cloud computing, system virtualization, SDN, and NFV to enhance resource management in data centers. In addition, Cloud is regarded as the foundation block for *Cloud Radio Access Network (CRAN)*, an emerging cellular framework that aims at meeting ever-growing end-users demand on 5G. In CRAN, the traditional base stations are split into radio and baseband parts. The radio part resides in the base station in the form of Remote Radio Head (RRH) unit and the baseband part in placed to Cloud for creating a centralized and virtualized Baseband Unit (BBU) pool for different base stations.

**Mobile Edge Computing (MEC):** Among the user proximate computing paradigms, Mobile Edge Computing (MEC) is considered as one of the key enablers of 5G. Unlike CRAN [48], in MEC, base stations and access points are equipped with Edge servers that take care of 5G related issues at the edge network. MEC facilitates a computationally enriched distributed RAN architecture upon the LTE-based networking. Ongoing researches on MEC targets real-time context awareness [49], dynamic computation offloading [50], energy efficiency [51] and multi-media caching [52] for 5G networking.

**Edge and Fog Computing:** Edge and Fog computing are coined to complement remote Cloud to meet the service demand of a geographically distributed large number of IoT devices. In Edge computing, the embedded computation capabilities of IoT devices or local resources accessed via ad-hoc networking are used to process IoT data. Usually, Edge computing paradigm is well suited to perform light computational tasks and does not probe global Internet unless intervention of remote (core) Cloud is required. However, not all the IoT devices are computationally enabled, or local Edge resources are computational-enriched to execute different large-scale IoT applications simultaneously. In this case, executing latency sensitive IoT applications at remote Cloud can degrade the QoS significantly [60]. Moreover, a huge amount of IoT workload sent to remote Cloud can flood the global internet and congest the network. Therefore, Fog computing is coined that offers infrastructure and software services through distributed Fog nodes to execute IoT applications within the network [54].

In Fog computing, traditional networking devices such as routers, switches, set-top boxes and proxy servers along with dedicated Nano-servers and Micro-datacenters can act as Fog nodes and create a wide area Cloud-like services both in independent or clustered manner [55]. Mobile Edge servers or Cloudlets [53] can also be regarded as Fog nodes to conduct their respective jobs in Fog enabled Mobile Cloud Computing and MEC. In some cases, Edge and Fog computing are used interchangeably although, in a broader perspective, Edge is considered as a subset of Fog Computing [56]. However, in Edge and Fog computing, the integration of 5G has already been discussed in terms of bandwidth management during computing instance migration [57] and SDN-enabled IoT resource discovery [58]. The concept of Fog radio access network (FRAN) [59] is also getting attention from both academia and industry where Fog resources are used to create BBU pool for the base stations.

Working principle of these computing paradigms largely depends on virtualization techniques. The alignment of 5G with different computing paradigms can also be analyzed through the interplay between network and resource virtualization techniques. Network Slicing is one of the key features of 5G network virtualization. Computing paradigms can also extend the vision of 5G network slicing into data center and Fog nodes. By the latter, we mean that the vision of network slicing can be applied to the shared data center network infrastructure and Fog networks to provide an end-to-end logical network for applications by establishing a full-stack virtualized environment. This form of network slicing can also be expanded beyond a data center networks into multi-Clouds or even cluster of Fog nodes [14]. Whatever the extension may be, this creates a new set of challenges to the network, including Wide Area Network (WAN) segments, cloud data centers (DCs) and Fog resources.

## 4.3  Network Slicing in 5G

In recent years, numerous research initiatives are taken by industries and academia to explore different aspects of 5G. Network architecture and its associated physical and MAC layer management are among the prime focuses of current 5G research works. The impact of 5G in different real-world applications, sustainability, and quality expectations are also getting predominant in the research arena. However, among the ongoing researches in 5G, network slicing is drawing more attractions since this distinctive feature of 5G aims at supporting diverse requirements at the finest granularity over a shared network infrastructure [6][7].

Network slicing in 5G refers to sharing a physical network's resources to multiple virtual networks. More precisely, network slices are regarded as a set of virtualized networks on the top of a physical network [8]. The network slices can be allocated to specific applications/services, use cases or business models to meet their requirements.

Each network slice can be operated independently with its own virtual resources, topology, data traffic flow, management policies, and protocols. Network slicing usually requires implementation in an end-to-end manner to support co-existence of heterogeneous systems [9].

The network slicing paves the way for customized connectivity among a high number of inter-connected end-to-end devices. It enhances network automation and leverages the full capacity of SDN and NFV. Also, it helps to make the traditional networking architecture scalable according to the context. Since network slicing shares a common underlying infrastructure to multiple virtualized networks, it is considered as one of the most cost-effective ways to use network resources and reduce both capital and operational expenses [10]. Besides, it ensures that the reliability and limitations (congestion, security issues) of one slice do not affect the others. Network slicing assists isolation and protection of data, control and management plane that enforce security within the network. Moreover, network slicing can be extended to multiple computing paradigms such as Edge [11], Fog [14] and Cloud that eventually improves their interoperability and helps to bring services closer to the end user with less Service Level Agreement (SLA) violations [12].

Apart from the benefits, the network slicing in current 5G context is subjected to diversified challenges, however. Resource provisioning among multiple virtual networks is difficult to achieve since each virtual network has a different level of resource affinity and it can be changed with the course of time. Besides, mobility management and wireless resource virtualization can intensify the network slicing problems in 5G. End-to-End slice orchestration and management can also make network slicing complicated.

Recent researches in 5G network slicing mainly focus on addressing the challenges through efficient network slicing frameworks. Extending the literature [12][13], we depicted a generic framework for 5G network slicing in Figure 4.1 The framework consists of three main layers: *Infrastructure layer*, *Network Function layer*, and *Service layer*.

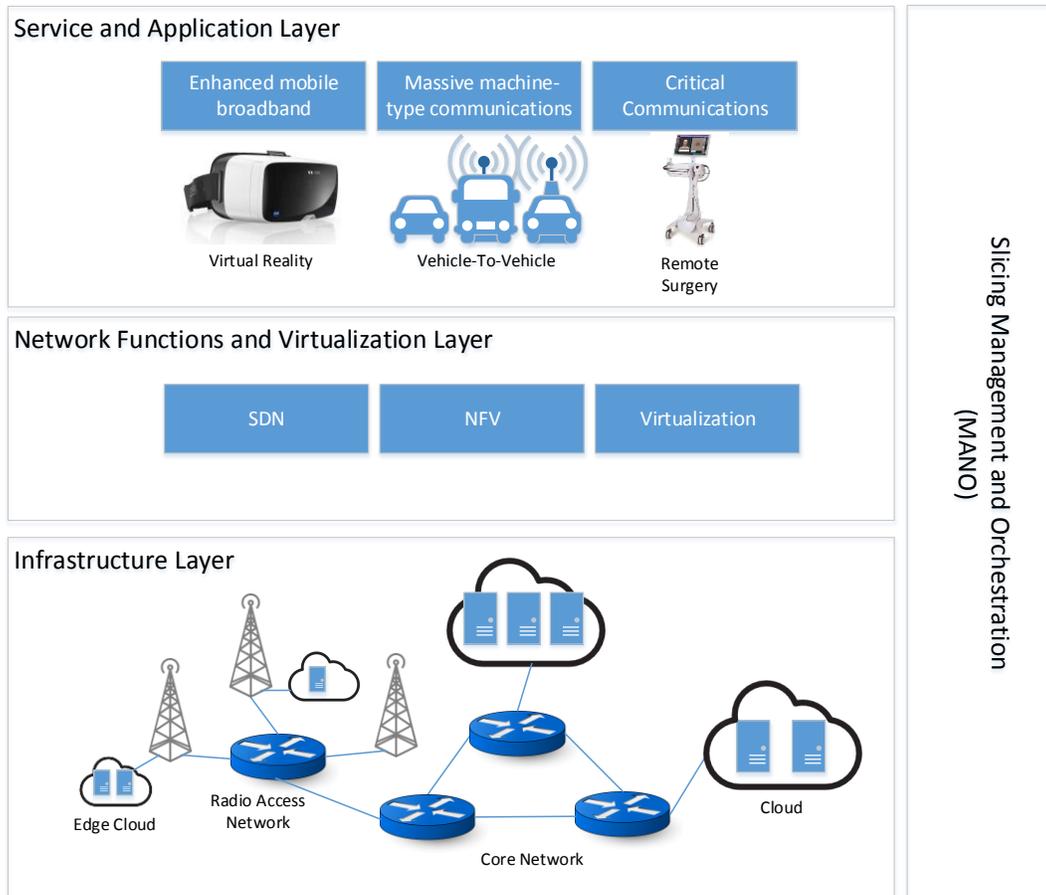

Figure 4.1: Generic 5G Slicing Framework.

**Infrastructure layer:** The infrastructure layer defines the actual physical network architecture. It can be expanded from Edge Cloud to remote Cloud through radio access network and the core network. Different software defined techniques are encapsulated to facilitate resource abstraction within the core network and the radio access network. Besides, in this layer, several policies are conducted to deploy, control, manage and orchestrate the underlying infrastructure. This layer allocates resources (compute, storage, bandwidth, etc.) to network slices in such way that upper layers can get access to handle them according to the context.

**Network Function and Virtualization Layer:** The network function and virtualization layer executes all the required operations to manage the virtual resources and network function's life cycle. It also facilitates optimal placement of network slices to virtual resources and chaining of multiple slices so that they can meet specific requirements of a particular service or application. SDN, NFV and different virtualization techniques are considered as the significant technical aspect of this layer. This layer explicitly manages the functionality of core and local radio access network. It can handle both coarse-grained and fine-grained network functions efficiently.

**Service and Application Layer:** The service and application layer can be composed by connected vehicles, virtual reality appliances, mobile devices, etc. having a specific use case or business model and represent certain utility expectations from the networking infrastructure and the network functions. Based on requirements or high-level description of the service or applications, virtualized network functions are mapped to physical resources in such way that SLA for the respective application or service does not get violated.

**Slicing Management and Orchestration (MANO):** The functionality of the above layers are explicitly monitored and managed by the slicing management and orchestration layer. The main task of this layer includes;

1. Creation of virtual network instances upon the physical network by using the functionality of the infrastructure layer.

2. Mapping of network functions to virtualized network instances to build a service chain with the association of network function and virtualization layer.

3. Maintaining communication between service/application and the network slicing framework to manage the lifecycle of virtual network instances and dynamically adapt or scale the virtualized resources according to the changing context.

The logical framework of 5G network slicing is still evolving. Retaining the basic structure, extension of this framework to handle the future dynamics of network slicing can be a potential approach to further standardization of 5G.

According to Huawei high-level perspective of 5G network [42], Cloud-Native network architecture for 5G has the following characteristics: 1) it provides Cloud data center based architecture and logically independent network slicing on the network infrastructure to support different application scenarios. 2) It uses Cloud-RAN[1] to build radio access networks (RAN) to provide a substantial number of connections and implement 5G required on-demand deployments of RAN functions. 3) It provides simpler core network architecture and provides on-demand configuration of network functions via user and control plane separation, unified database management, and component-based functions, and. 4) In automatic manner, it implements network slicing service to reduce operating expenses.

In the following section, we intend to review the state-of-the-art related work on network slice management happening in Cloud computing literature. Our survey in this area can help researcher to apply advances and innovation in 5G and Clouds reciprocally.

---

[1] CLOUD-RAN (CRAN) is a centralized architecture for radio access network (RAN) in which the radio transceivers are separated from the digital baseband processors. This means that operators can centralize multiple base band units in one location. This simplifies the amount of equipment needed at each individual cell site. Ultimately, the network functions in this architecture become virtualized in the Cloud.

13## 4.4 Network Slicing in Software Defined Clouds

Virtualization technology has been the cornerstone of the resource management and optimization in Cloud data centers for the last decade. Many research proposals have been expressed for VM placement and Virtual Machine (VM) migration to improve utilization and efficiency of both physical and virtual servers [15]. In this section, we focus on the state of the art network-aware VM/VNF management in line with the aim of the report, i.e., network slicing management for SDCs. Figure 4.2 illustrates our proposed taxonomy of network-aware VM/VNF management in SDCS. Our taxonomy classifies existing works based on the objective of the research, the approach used to address the problem, the exploited optimization technique, and finally the evaluation technique used to validate the approach. In the remaining parts of this section, we cover network slicing from three different perspectives and map them to the proposed taxonomy: Network-aware VM management, Network-aware VM migration, and VNF management.

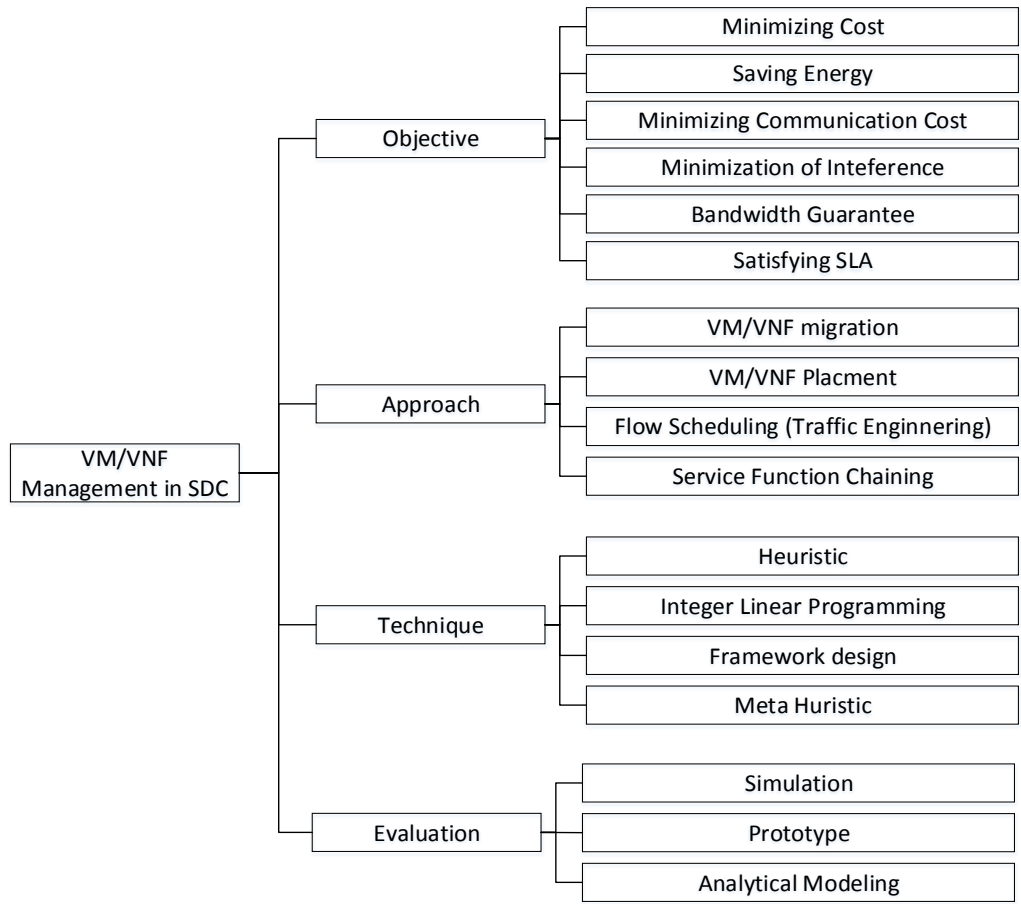

Figure 4.2: Taxonomy of network-aware VM/VNF Management in software-defined Clouds

## 4.4.1 Network-aware Virtual Machines Management

Cziva et al. [15] present an orchestration framework to exploit time-based network information to live migrate VMs and minimize the network cost. Wang et al. [16] propose a VM placement mechanism to reduce the number of hops between communicating VMs, save energy, and balance the network load. Remedy [17] relies on SDN to monitor the state of the network and estimate the cost of VM migration. Their technique detects congested links and migrates VMs to remove congestion on those links.

Jiang et al. [18] worked on joint VM placement and network routing problem of data centers to minimize network cost in real-time. They proposed an online algorithm to optimize the VM placement and data traffic routing with dynamically adapting traffic loads. VMPlanner [19] also optimizes VM placement and network routing. The solution includes VM grouping that consolidates VMs with high inter-group traffic, VM group placement within a rack, and traffic consolidation to minimize the rack traffic. Jin et al. [21] studied joint host-network optimization problem. The problem is formulated as an integer linear problem which combines VM placement and routing problem. Cui et al. [20] explore the joint policy-aware and network-aware VM migration problem and present a VM management to reduce network-wide communication cost in data center networks while considering the policies regarding the network functions and middleboxes. Table 4.2 summarizes the research projects on network-aware VM management.

16Table 4.2 - Network-aware Virtual Machines Management

| Project | Objectives | Approach/Technique | Evaluation |
|---|---|---|---|
| Cziva et al. [15] | Minimization of the network communication cost | VM migration – Framework Design | Prototype |
| Wang et al. [16] | Reducing the number of hops between communicating VMs and network power consumption | VM placement – Heuristic | Simulation |
| Remedy [17] | Removing congestion in the network | VM migration – Framework Design | Simulation |
| Jiang et al. [18] | Minimization of the network communication cost | VM Placement and Migration – Heuristic (Markov approximation) | Simulation |
| VMPlanner [19] | Reducing network power consumption | VM placement and traffic flow routing - Heuristic | Simulation |
| PLAN [20] | Minimization of the network communication cost while meeting network policy requirements | VM Placement - Heuristic | Prototype/Simulation |

## *4.4.2 Network-aware Virtual Machine Migration Planning*

A large body of literature focused on improving the efficiency of VM migration mechanism [22]. Bari et al. [23] propose a method for finding an efficient migration plan. They try to find a sequence of migrations to move a group of VMs to their final destinations while migration time is minimized. In their method, they monitor residual bandwidth available on the links between source and destination after performing each step in the sequence. Similarly, Ghorbani et al. [24] propose an algorithm to generate an ordered list of VMs to migrate and a set of forwarding flow changes. They concentrate on imposing bandwidth guarantees on the links to ensure that link capacity is not violated during the migration. The VM migration planning problem is also tackled by Li et al. [25] where they address the workload-aware migration problem and propose methods for selection of candidate virtual machines, destination hosts, and sequence for migration. All these studies focus on the migration order of a group of VMs while taking into account network cost. Xu et al. [26] propose an interference-aware VM live migration plan called *iAware* that minimizes both migration and co-location interference among VMs. Table 4.3 summarizes the research projects on VM migration planning.

Table 4.3 - Virtual Machine Migration Planning

| Project | Objectives | Approach/Technique | Evaluation |
|---|---|---|---|
| Bari et al. [23] | Finding sequence of migrations to while migration time is minimized | VM migration – Heuristic | Simulation |
| Ghorbani et al. [24] | Finding sequence of migrations while imposing bandwidth guarantees | VM migration – Heuristic | Simulation |
| Li et al. [25] | Finding sequence of migrations and destination hosts to balance the load | VM migration – Heuristic | Simulation |
| iAware [26] | Minimization of migration and co-location interference among VMs | VM migration – Heuristic | Prototype/Simulation |

## *4.4.3 Virtual Network Functions Management*

Network Functions Virtualization (NFV) is an emerging paradigm where network functions such as firewalls, Network Address Translation (NAT), Virtual Private Network (VPN), etc. are virtualized and divided up into multiple building blocks called Virtualized Network Functions (VNFs). VNFs are often chained together and build Service Function Chains (SFC) to deliver a required network functionality. Han et al. [27] present a comprehensive survey of key challenges and technical requirements of NFV where they present an architectural framework for NFV. They focus on the efficient instantiation, placement and migration of VNFs and network performance. VNF-P is a model proposed by Moens and Turck [28] for efficient placement of VNFs. They propose a NFV burst scenario in a hybrid scenario in which the base demand for network function service is handled by physical resources while the extra load is handled by virtual service instances. Cloud4NFV [29] is a platform following the NFV standards by European Telecommunications Standards Institute (ETSI) to build Network Function as a Service using a Cloud platform. Their VNF Orchestrator exposes RESTful APIs allowing VNF deployment. A Cloud platform such as OpenStack supports management of virtual infrastructure at the background. vConductor [30] is another NFV management system proposed by Shen et al. for the end-to-end virtual network services. vConductor has simple graphical user interfaces (GUIs) for automatic provisioning of virtual network services and supports the management of VNFs and existing physical network functions. MORSA [31] proposed as part of vConductor to perform virtual machine (VM) placement for building NFV infrastructure in the presence of conflicting objectives of involving stakeholders such as users, Cloud providers, and telecommunication network operators.

Service chain is a series of VMs hosting VNFs in a designated order with a flow goes through them sequentially to provide desired network functionality. Tabular VM migration (TVM) proposed by [32] aims at reducing the number of hops in service chain of network functions in Cloud data centers. They use VM migration to reduce the number of hops (network elements) the flow should traverse to satisfy Service level agreements (SLAs). SLA-driven Ordered Variable-width Windowing (SOVWin) is a heuristic proposed by Pai et al. [33] to address the same problem, however, using initial static placement. Similarly, an orchestrator for the automated placement of VNFs across the resources proposed by Clayman et al. [34].

Table 4.4 - Virtual Network Functions Management Projects

| Project | Objectives | Approach/Technique |
|---|---|---|
| VNF-P | Handling burst in network services demand while minimizing the number of servers | Resource Allocation - Integer linear programming (ILP) |
| Cloud4NFV | Providing Network Function as a Service | Service provisioning – Framework Design |
| vConductor | Virtual network services provisioning and management | Service provisioning – Framework Design |
| MORSA | Multi Objective placement of virtual services | Placement - Multi-objective Genetic Algorithm |
| TVM | Reducing number of hops in service chain | VNF Migration - Heuristic |
| SOVWin | Increasing user requests acceptance rate and minimization of SLA violation | VNF Placement - Heuristic |
| Clayman et al. | Providing automatic placement of the virtual nodes | VNF Placement - Heuristic |
| T-NOVA | Building a Marketplace for VNF | Marketplace – Framework Design |
| UNIFY | Automated, dynamic service creation and service function chaining | Service provisioning– Framework Design |

The EU-funded T-NOVA project [35] aims to realize the NFaaS concept. It designs and implements integrated management and orchestrator platform for the automated provisioning, management, monitoring and optimization of VNFs. UNIFY [36] is another EU-funded FP7 project aims at supporting automated, dynamic service creation based on a fine-granular SFC model, SDN, and Cloud virtualization techniques. For more details on SFC, interested readers are referred to the literature survey by Medhat et al. [37]. Table 4.4 summarizes the state of the art projects on VNF management.

## 4.5 Network Slicing Management in Edge and Fog

Fog computing is a new trend in Cloud computing that intends to address the quality of service requirements of applications requiring real-time and low latency processing. While Fog acts as a middle layer between Edge and core Clouds to serve applications close to the data source, core Cloud data centers provide massive data storage, heavy-duty computation, or widearea connectivity for the application.

One of the key visions of Fog computing is to add compute capabilities or general purpose computing to Edge network devices such as mobile base stations, gateways and routers. On the other hand, SDN and NFV play key roles in prospective solutions to facilitate efficient management and orchestration of network services. Despite natural synergy and affinity between these technologies, there exist not many research on the integration of Fog/Edge computing and SDN/NFV as both are still in their infancy. In our view, intraction between SDN/NFV and Fog/Edge computing is crucial for emerging applications in IoT, 5G and stream analytics. However, the scope and requirements of such interaction is still an open problem. In the following, we provide an overview of the state-of-the-art within this context.

Lingen et al. [45] define a model-driven and service-centric architecture that addresses technical challenges of integrating NFV, Fog and 5G/MEC. They introduce an open architecture based on NFV MANO proposed by the European Telecommunications Standards Institute (ETSI) and aligned with the OpenFog Consortium (OFC) reference architecture[2] that offers uniform management of IoT services spanning through Cloud to the Edge. A two-layer abstraction model along with IoT-specific modules and enhanced NFV MANO architecture is proposed to integerate Cloud, network, and Fog. As a pilot study, they presented two use cases for physical security of Fog nodes and sensor telemetry through street cabinets in the city of Barcelona.

Truong et al. [43] are among the earliest who have proposed an SDN-based architecture to support Fog Computing. They have identified required components and specified their roles in the system. They also showed how their system can provide services in the context of Vehicular Adhoc Networks (VANETs). They showed benefits of their proposed architecture using two use-cases in data streaming and lane-change assistance services. In their proposed architecture, the centeral network view by the SDN Controller is utilized to manage resources and services and optimize their migration and replication.

---

[2] OpenFog Consortium, https://www.openfogconsortium.org/

Bruschi et al. [44] propose network slicing scheme for supporting multi-domain Fog/Cloud services. They propose SDN-based network slicing scheme to build an overlay network for geographically distributed Internet services using non-overlapping OpenFlow rules. Their experimental results show that the number of unicast forwarding rules installed in the overlay network significantly drops compared to the fully-meshed and OpenStack cases.

Inspired by Open Network Operating System (ONOS)[3] SDN controller, Choi et al. [46] propose a Fog operating system architecture called *FogOS* for IoT services. They identified four main challenges of Fog computing as: 1) *scalability* for handling significant number of IoT devices, 2) *complex inter-networking* caused by diverse forms of connectivity, e.g., various radio access technologies, 3) *dynamics and adaptation* in topology and quality of service (QoS) requirements, and finally 4) *diversity and heterogeneity* in communications, sensors, storage, and computing powers, etc. Based on these challenges, their proposed architecture consists of four main components: 1) Service and device abstraction, 2) Resource management, 3) Application management, 4) Edge resource: registration, ID/addressing, and control interface. They also demonstrate a preliminary proof-of-concept demonstration of their system for a drone-based surveillance service.

---

[3] ONOS, https://onosproject.org/

In a recent work, Diro et al. [47] propose a mixed SDN and Fog architecture which gives priority to critical network flows while takes into account fairness among other flows in the Fog-to-things communication to satisfy QoS requirements of heterogeneous IoT applications. They intend to satisfy QoS and perfromance measures such as packet delay, lost packets and maximize throughput. Results show that their proposed method is able to serve critical and urgent flows more efficiently while provides allocation of network slices to other flow classes.

## 4.6 Future Research Directions

In this section, we discuss open issues in software-defined Clouds and Edge computing environments along future directions.

### *4.6.1 Software Defined Clouds*

Our survey on network slicing management and orchestration in SDC shows that community very well recognized the problem of joint provisioning of hosts and network resources. In the earlier research, a vast amount of attention has been given to solutions for the optimization of cost/energy only focusing on either host [38] or network [39], not both. However, it is essential for the management component of the system to take into account both network and host cost at the same time. Otherwise, optimization of one can exacerbate the situation for the other. To address this issue, many research proposals have also focused on the joint host and network resource management. However, most of the proposed approaches suffer from high computational complexity, or they are not optimal. Therefore, the development of algorithms that manage joint hosts and network resource provisioning and scheduling is of great interest. In joint host and network resource management and orchestration, not only finding the minimum subset of hosts and network resources that can handle a given workload is crucial, but also SLA and users' QoS requirements (e.g., latency) must be satisfied. The problem of joint host and network resource provisioning becomes more sophisticated when SDC supports VNF and SFC.

SFC is a hot topic attainting a significant amount of attention by the community. However, little attention has been paid to VNF placement while meeting the QoS requirements of the applications. PLAN [20] intends to minimize the network communication cost while meeting network policy requirements. However, it only considers traditional middleboxes, and it does not take into account the option of VNF migration. Therefore, one of the areas requires more attention and development of novel optimization techniques is the management and orchestration of SFCs. This has to be done in a way that the placement and migration of VNFs are optimized while SLA violation and cost/energy are maximized.

Network-aware virtual machines management is a well-studied area. However, the majority of works in this context consider VM migration and VM placement to optimize network costs. The traffic engineering and dynamic flow scheduling combined with migration and placement of VMs also provide a promising direction for the minimization of network communication cost. For example, using SDN, management and orchestration module of the system can install flow entries on the switches of the shortest path with the lowest utilization to redirect VM migration traffic to an appropriate path.

The analytical modeling of SDCs has not been investigated intensely in the literature. Therefore building a model based on priority networks that can be used for analysis of the SDCs network and validation of results from experiments conducted via simulation.

Auto-scaling of VNFs is another area that requires more in-depth attention by the community. VNFs providing networking functions for the applications are subject to performance variation due to different factors such as the load of the service or overloaded underlying hosts. Therefore, development of auto-scaling mechanisms that monitor the performance of the VMs hosting VNFs and adaptively adds or remove VMs to satisfy the SLA requirements of the applications is of paramount importance for management and orchestration of network slices. In fact, efficient placement of VNFs [41] on hosts near to the service component producing data streams or users generating requests minimizes latency and reduces the overall network cost. However, placing it on a more powerful node far in the network improves processing time [40]. Existing solutions mostly focus on either scaling without placement or placement without scaling. Moreover, auto-scaling techniques of VNFs, they typically focus on auto-scaling of a single network service (e.g., firewall), while in practice auto-scaling of VNFs must be performed in accordance with SFCs. In this context, node and link capacity limits must be considered, and the solution must maximize the benefit gained from existing hardware using techniques such as dynamic pathing. Therefore, one of the promising avenues for future research on auto-scaling of VNFs is to explore the optimal dynamic resource allocation and placement.

## *4.6.2 Edge and Fog Computing*

In both Edge and Fog computing, the integration of 5G so far has been discussed within a very narrow scope. Although 5G network resource management and resource discovery in Edge/Fog computing have been investigated, many other challenging issues in this area are still unexplored. Mobility-aware service management in 5G enabled Fog computing and forwarding large amount of data from one Fog node to another in real-time overcoming communication overhead can be very difficult to ensure. In addition, due to decentralized orchestration and heterogeneity among Fog nodes, modelling, management and provisioning of 5G network resources are not as straight-forward as other computing paradigms.

Moreover, compared to Mobile Edge servers, Cloudlets and Cloud datacenters, the number of Fog nodes and their probability of being faulty are very high. In this case, implementation of SDN (one of the foundation blocks of 5G) in Fog computing can get obstructed significantly. One the other hand, Fog computing enables traditional networking devices to process incoming data and due to 5G, this data amount can be significantly huge. In such scenario, adding more resources in traditional networking devices will be very costly, less secured and hinders their inherent functionalities like routing, packet forwarding, etc. which in consequence affect the basic commitments of 5G network and NFV.

Nonetheless, Fog infrastructures can be owned by different providers that can significantly resist developing a generalized pricing policy for 5G-enabled Fog computing. Prioritized network slicing for forwarding latency-sensitive IoT data can also contribute additional complications in 5G enabled Fog computing. Opportunistic scheduling and reservation of virtual network resources is tough to implement in Fog as it deals with a large number of IoT devices and their data sensing frequency can change with the course of time. Balancing load on different virtual networks and their QoS can degrade significantly unless efficient monitoring is imposed. Since Fog computing is a distributed computing paradigm, centralized monitoring of network resources can intensify the problem. In this case, distributed monitoring can be an efficient solution, although it can be failed to reflect the whole network context in a body. Extensive research is required to solve this issue. Besides, in promoting fault-tolerance of 5G-enabled Fog computing, topology-aware application placement, dynamic fault detection and reactive management can play a significant role which is subjected to uneven characteristics of the Fog nodes.

## 4.7 Conclusion

In this paper, we intended to investigate research proposals for the management and orchestration of network slices in different platforms. We discussed emerging technologies such as Software-defined networking SDN and NFV. We explored the vision of 5G for network slicing and discussed some of the ongoing projects and studies in this area. We surveyed the state of the art approaches to network slicing in Software-defined Clouds and application of this vision to the Cloud computing context. We disscussed the state of the art literature on network slices in emerging Fog/Edge computing. Finally, we identified gaps in this context and provided future directions towards the notion of network slicing.

# Acknowledgments

This work is supported through Huawei Innovation Research Program (HIRP). We also thank Zhouwei for his comments and support for the work.